\documentclass{aa}
\usepackage{graphicx}
\input{psfig.sty}
\begin{document}
   \title{Structure and mass distribution of spiral galaxies at 
   intermediate redshifts}
   \author{A. Tamm\inst{1} \and
           P. Tenjes\inst{1, 2} }
   \offprints{P. Tenjes}
   \institute{Institute of Theoretical Physics, Tartu University,
              T\"ahe 4, Tartu, 51010 Estonia\\
              \email{atamm@ut.ee; ptenjes@ut.ee}
         \and
             Tartu Observatory, T\~oravere, Tartumaa,
             61602 Estonia}
   \date{Received ..... ; accepted .............}

   \abstract{Using the HST archive WFPC2 observations and rotation curves 
measured by Vogt et al. (\cite{vogt1}), we constructed self-consistent 
light and mass distribution models for three disk galaxies at redshifts 
$z=0.15, 0.90$ and $0.99$. The models consist of three components: the bulge, 
the disk and the dark matter. Spatial density distribution parameters
for the components were calculated. After applying k-corrections, 
mass-to-light ratios for galactic disks within the maximum disk assumption 
are $M/L_B =$ 4.4, 1.2 and 1.2, respectively. Corresponding central densities 
of dark matter halos within a truncated isothermal model are 0.0092, 0.028 and 
0.015 in units $\rm M_{\sun} /pc^3$. The light distribution of the
galaxies in outer parts is steeper than a simple exponential disk.
   }
\authorrunning{Tamm \& Tenjes}
\titlerunning{spiral galaxies at intermediate redshifts}
   \maketitle
%

\section{Introduction}

The study of dark matter halo properties and central densities allows 
to constrain possible galaxy formation models and large scale stucture 
formation scenarios (Primack \cite{primack}; Khairul Alam et al. \cite{khai}). 
It is interesting to analyse the corresponding data at different 
redshifts. For this kind of analysis it is needed to know both the 
distribution of visible and dark matter. Unfortunately, the structure 
and mass distribution of stellar populations, and therefore, the 
discrimination between visible and dark matter in very distant galaxies 
is still largely unknown. 

General morphological evolution of galaxies at intermediate and 
high redshifts has been studied, for example by Smail et al. 
(\cite{smai}), Andreon et al. (\cite{andr}), Lubin et al. (\cite{lubi}), 
Fabricant et al. (\cite{fabr}), van den Bergh et al. (\cite{berg}), 
van Dokkum et al. (\cite{dokk3}), La Barbera et al. (\cite{barb}), 
Im et al. (\cite{im}). Several statistical correlations and color 
gradients in clusters at redshifts $z = 0.3-1$ have been studied  by 
Stanford et al. (\cite{stan}), Ziegler et al. (\cite{zieg1}), van 
Dokkum et al. (\cite{dokk2}), Saglia et al. (\cite{sagl}), Tamura 
\& Ohta (\cite{ta:oh}). More sophisticated correlation parameters 
for elliptical galaxies, the Fundamental Plane parameters involving 
dynamical information, have been determined by Kelson et al. 
(\cite{kels1}), van Dokkum et al. (\cite{dokk1}, \cite{dokk4}), Kelson 
et al. (\cite{kels2}), Treu et al. (\cite{treu}).

These studies indicate that most of the stellar content of early-type 
galaxies has formed at redshifts $z > 2-3$, and has passed normal chemical
evolution and there is no significant age difference between field 
and cluster galaxies. Within the standard $\Lambda$CDM model, field ellipticals
must be younger than cluster ellipticals. On the other hand, the fraction 
of early-type galaxies in clusters is smaller and the fraction of E+A
(post-starburst) galaxies is higher at intermediate redshifts, indicating 
some morphological transformations from $z\sim 1 $ to the present time. 
Theoretical models indicate that truncation of star formation rate can 
successfully explain this kind of morphological changes (van Dokkum \& Franx, 
\cite{do:fr2}; Bicker et al. \cite{bick}). 

Less is known about the structure of spiral galaxies at intermediate 
redshifts. For analysing the emission line profiles of disk
galaxies, it is needed to distinguish the rotation and dispersion  
components. Spectrograph slit width is usually 
only slightly smaller than galactic dimensions which makes derived 
rotation curves uncertain and often only maximum values of rotation
velocities can be estimated.

By now the maximum values of rotation velocities have been measured for
distant galaxies up to redshift $z = 1.34$ (Colless et al. \cite{coll}; 
Forbes et al. \cite{forb}; Rix et al. \cite{rix}; Simard \& Pritchet
\cite{si:pr}, Schade et al. \cite{scha}; Mallen-Ornelas et al. 
\cite{mall}; Briggs et al. \cite{brig}; van Dokkum \& Stanford 
\cite{do:st}). When plotted to the Tully-Fisher diagram, these measurements 
allow to compare the measured galaxies with the local ones. Rotation curves 
of very distant galaxies have been measured only by Vogt et al. 
(\cite{vogt1}, \cite{vogt2}), Ziegler et al. (\cite{zieg2}) and Steidel
et al. (\cite{stei}). These data allow more detailed modeling of the structure
of corresponding galaxies, especially if the curve extends  
beyond the solid-body rotation part.  

To construct a self-consistent photometrical and dynamical model of a
galaxy, both surface photometry and rotation curve data are needed. In the
case of galaxies at intermediate redshifts the typical scale is $5-10$~kpc 
per arcsec and in order to determine the parameters of galactic
components, it is recommendable to use also high-resolution HST photometry
in addition to other photometrical data.

For our modeling, the galaxies had to match the following criteria. 
Their velocity profiles had to be undisturbed, symmetric with respect 
to the galactic centre and with sufficient extent. Galaxies seen 
almost face-on were not suitable due to uncertainties of the inclination 
angle of the galactic plane. We also could not use galaxies being seen 
edge-on, because for these galaxies the surface brightness distribution is
largely influenced by galactic absorption. For detailed modeling, we
selected three galaxies with rotation curves measured by Vogt et al.
(\cite{vogt1}): 074-2237, 064-4412, 094-2210. For deriving the surface
brightness distribution of these galaxies, we used images from the HST archive.
General properties of these galaxies are given in Table~\ref{tab1}. In the
present work we take $H_0 = 65$~km/s/Mpc and $q_0 = 0.5.$

\begin{table*}
   \caption[]{General galactic parameters.}
   \label{tab1}
\begin{tabular}{llllllll}
\hline
Name$^a$ &  RA          & DEC         &Hubble&$z^a$& Scale &Inclin.$^a$&$M_B^b$\\
         &  (2000)      & (2000)      & type &     & (kpc/'')&(deg)&      \\
\hline
074-2237 &$\rm 14^h17^m45.4^s$&$\rm 52^o27'59.9''$& Sc  & 0.154 & 2.76 & 80 &  $-20.0$\\
064-4412 &$\rm 14^h17^m53.4^s$&$\rm 52^o29'41.1''$& Sc  & 0.988 & 6.71 & 68 &  $-20.6$\\
094-2210 &$\rm 14^h17^m31.3^s$&$\rm 52^o26'08.8''$& Sbc & 0.900 & 6.65 & 60 &  $-21.5$\\
\hline
\end{tabular}
\begin{list}{}{}
\item[$^a$] Galactic names and redshifts are from Vogt et al.
(\cite{vogt1}), inclinations are from Simard et al. (\cite{sima})\\
\item[$^b$] Absolute B-magnitudes in galactic rest-frame are from
our models (Sect.~4)\\
\end{list}
\end{table*}

\section{Photometrical structure}

The images of all three galaxies were retrieved from the archive of 
the Hubble Space Telescope (WFPC2, observational programs 5090 
and 5109, filters F814W and F606W). For processing we used 
pipeline-reduced on-the-flight calibrated images. 12 exposures in both 
colors were available for the galaxy 074-2237 and 4 exposures for 
the other two. Details about image processing and derivation of
surface brihtness profiles are described in Tamm \& Tenjes 
(\cite{ta:te}). Here we summarize only basic steps, bringing out
especially these aspects which differ from the previous processing.
Processing was done with IRAF/STSDAS packages.

\begin{figure*}
\begin{center}
\setlength{\unitlength}{1cm}
\begin{picture}(18,6)
\put(0.0,-1.0){\includegraphics{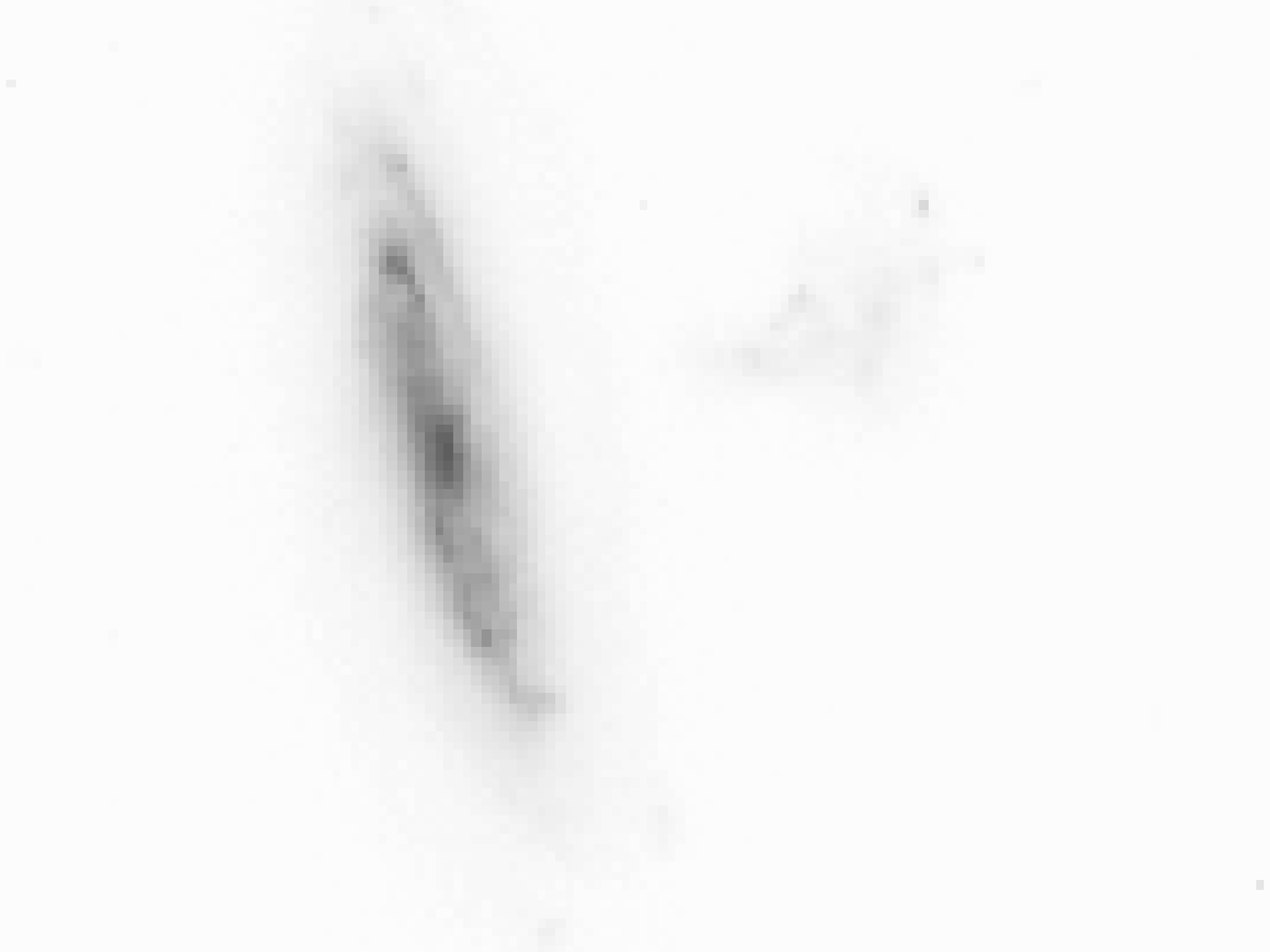}}
\put(6.0,-1.0){\includegraphics{g2.eps}}
\put(12.0,-1.0){\includegraphics{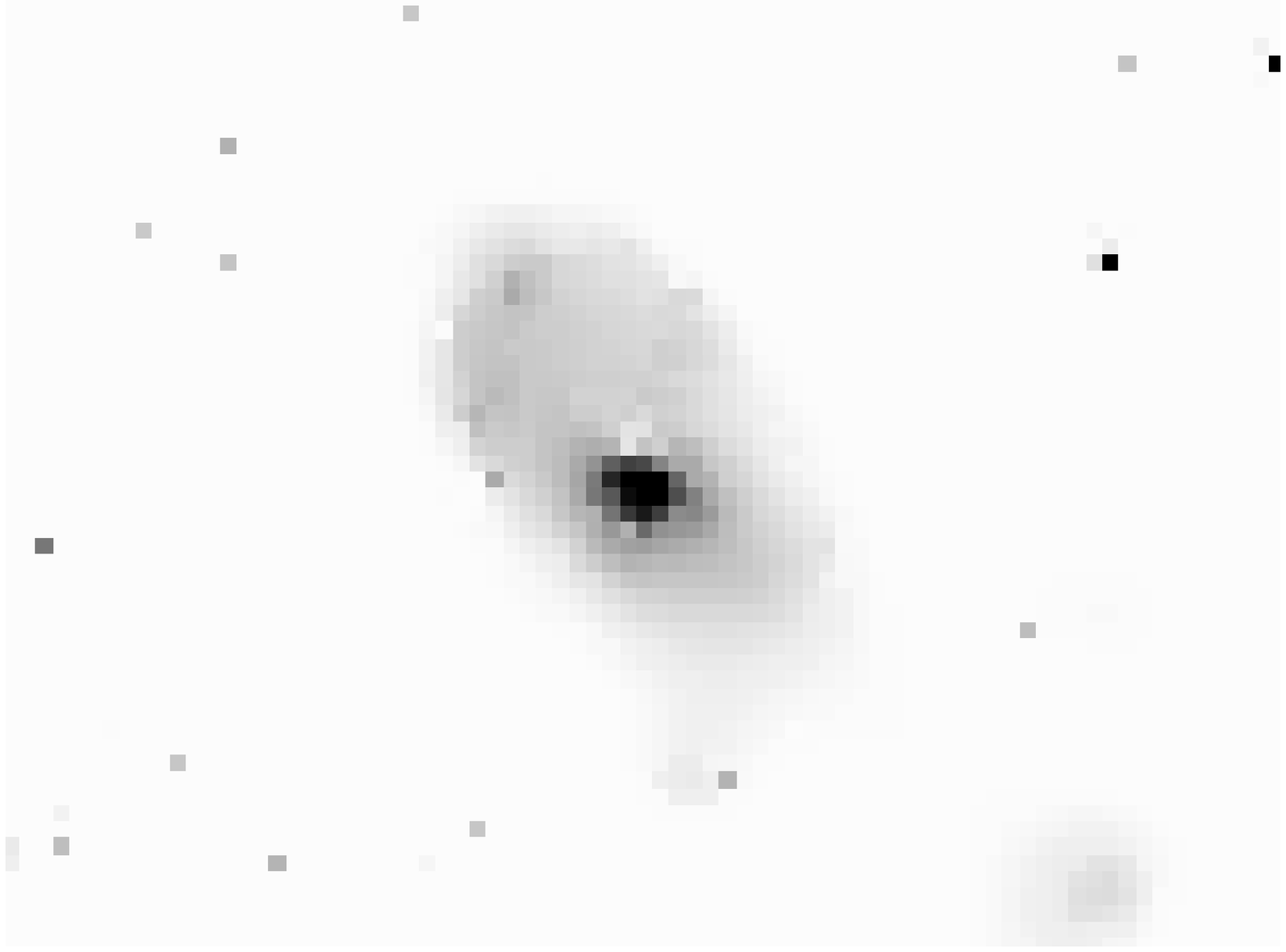}}
\end{picture}
\end{center}
\caption{Images of the galaxies (from left to right) 074-2237, 064-4412, 
         and 094-2210 in F814W color after background subtraction.}
\label{g123}
\end{figure*}

In order to elliminate cosmic rays, exposures were combined using 
the STSDAS task {\em combine} (option {\em crrej}). Before estimating 
the background, combined images were filtered to smooth down 
noises (task {\em adaptive}). We found no evidence of galaxy distortion
caused by adaptive filtering. As the investigated galaxies  
were rather small in CCD frames, it was sufficient to subtract in all 
cases a constant background value from the images. Background values 
were determined by interpolating the values of the surrounding regions
of the galaxies. This stage differs from the one we have used earlier. New
on-the-flight calibration of the HST archive data has improved the quality
of the original images, especially in the case of the galaxy 064-4412. The
background levels were (in WFPC2 intensity counts) 11.7 (I)and 19.8 (V) for
the galaxy 074-2262; 6.40 (I) and 7.25 (V) for the galaxy 064-4412; 6.25
(I) and 7.25 (V) for the galaxy 094-2210. Images of the three galaxies
after background subtraction are given in Fig.~\ref{g123}.

\begin{figure*}
\begin{center}
\setlength{\unitlength}{1cm}
\begin{picture}(18,6)
\put(-0.5,5.5){\includegraphics{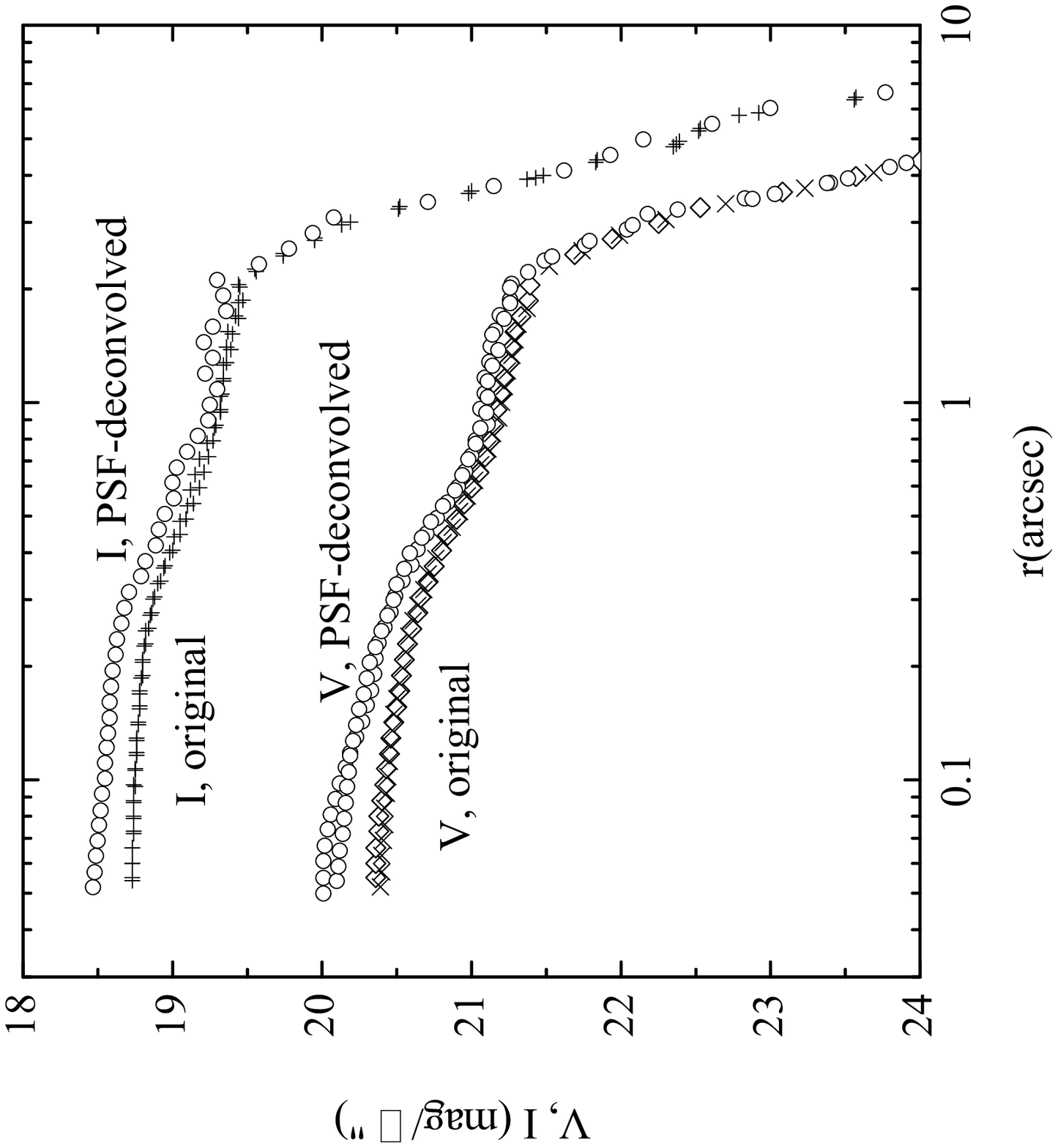}}
\put(5.4,5.5){\includegraphics{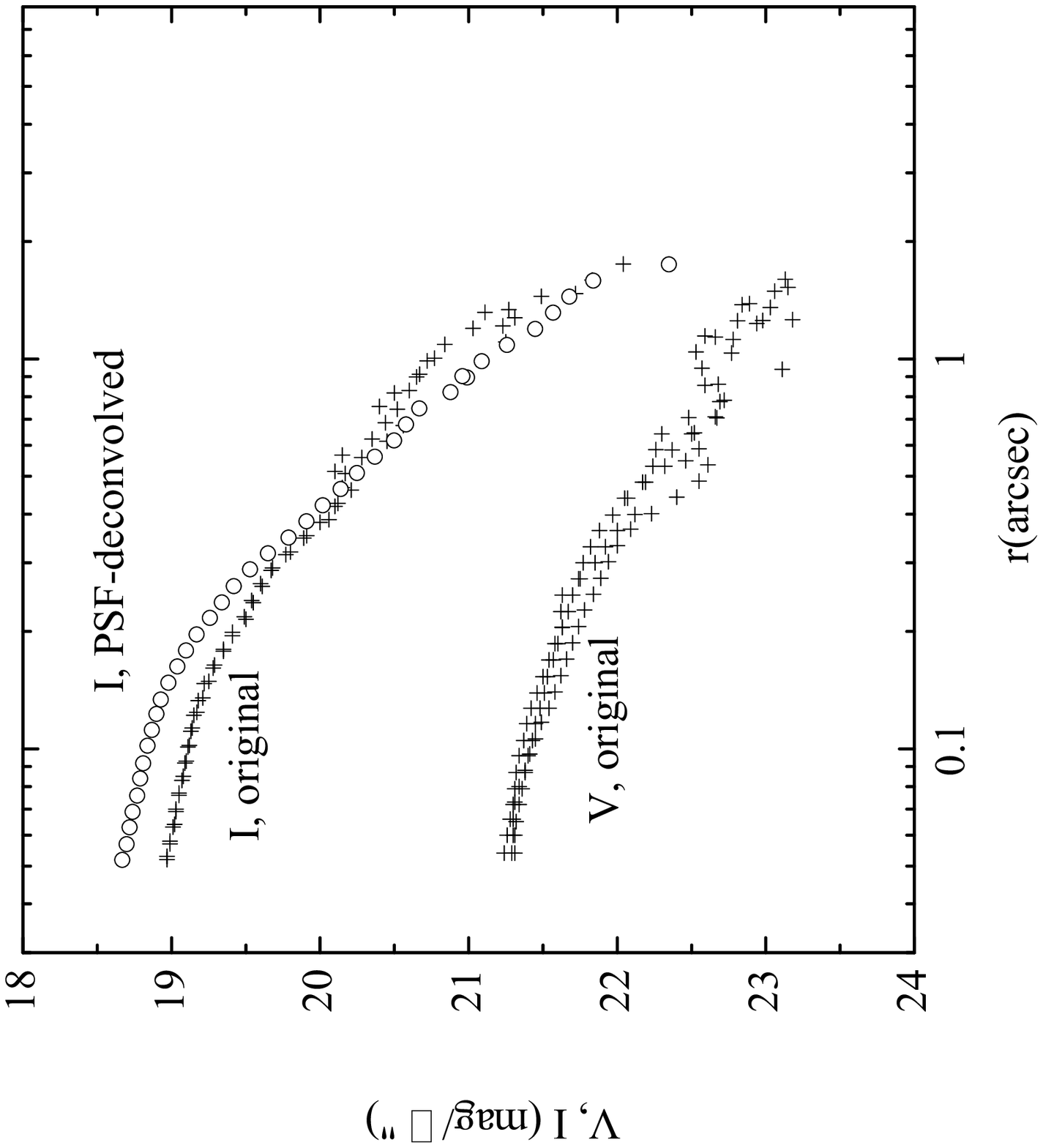}}
\put(11.3,5.5){\includegraphics{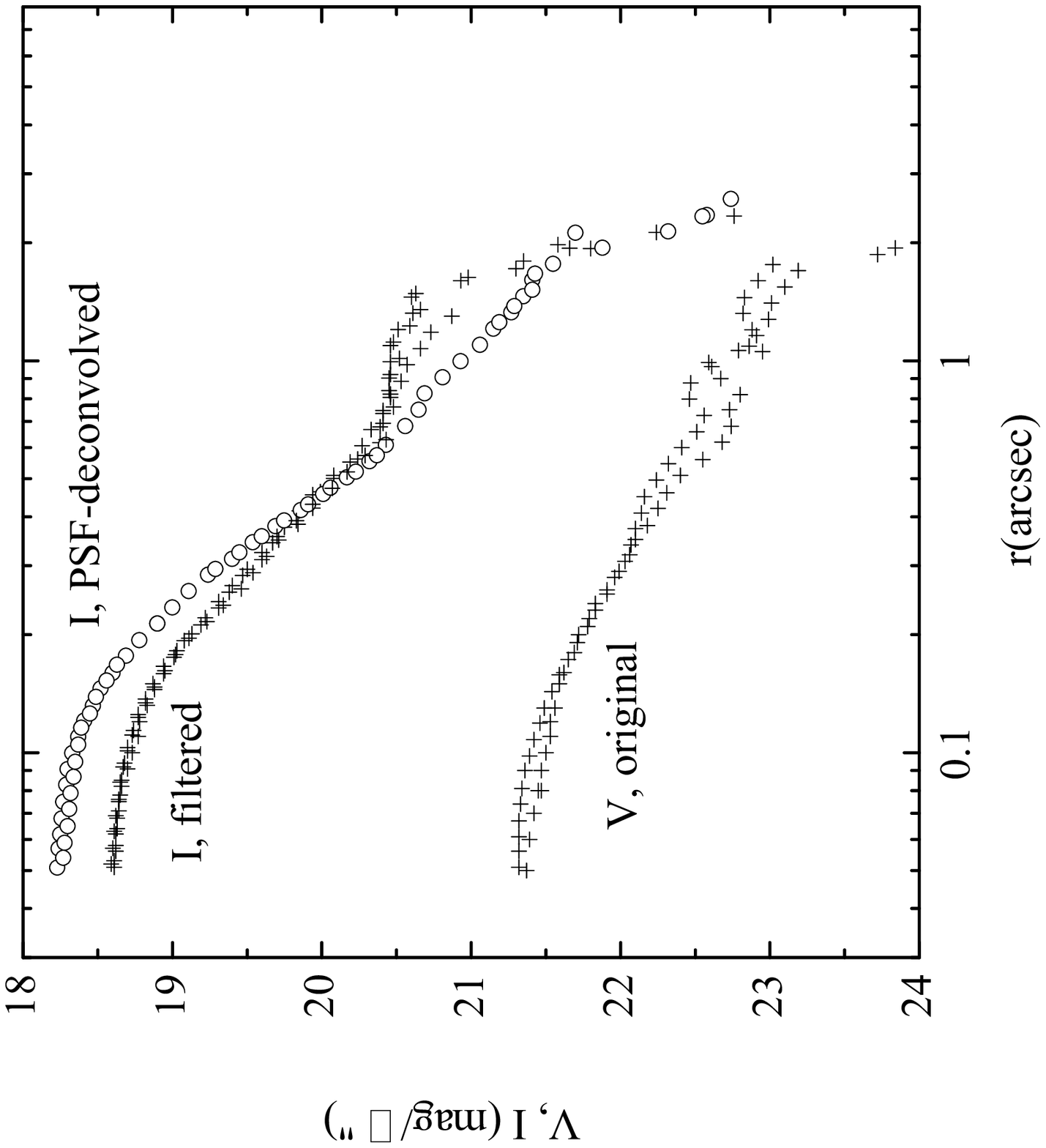}}
\end{picture}
\end{center}
\caption{Surface brightness distribution of the galaxies (from left to 
         right) 074-2237, 064-4412, and 094-2210 in I and V colors. Both 
         original and deconvolved profiles are given.}
\label{g123iv}
\end{figure*}

The resulting images were deconvolved using the Lucy-Richardson algorithm for
modeling PSF with the help of TinyTim 6.0. V-band images of 064-4412 and
094-2210 were not deconvolved, because due to a low signal-to-noise ratio
deconvolution process gave unrealistic results. In order to calculate 
isophotes, task {\em ellipse} was used. The fitting process was carried out 
several times for each of the galaxies in order to estimate and reduce 
uncertainties. The scatter of the dots in Fig.~\ref{g123iv} illustrates
uncertainties of isophote fitting. 

Before final calibration, we estimated the corrections of the
luminosity flux for the geometric distortion and the charge transfer
efficiency effect of the WFPC2 (Holtzman et al. \cite{holt1}). We added 
2\% to the signal to compensate for the geometric distortion of 064-4412 
and 094-2210 and 1.5\% in the case of 074-2237. To transfer the 
luminosity counts into standard Johnson magnitudes, the calibration 
described by Holtzman et al. (\cite{holt2}) (H95) was used. For the 
filter F814W, we used the flight photometric system Eq.~(8) and the coefficients 
from Table 7 in H95:
$$I = -2.5~\log ~(DN/t) - 0.062(V-I) + 0.025(V-I)^2 + $$
$$20.839 + 2.5\log(GR) + 0.1 + 0.05.$$
For F606W filter, we used the synthetic calibration Eq.~(9) and
coefficients from Table 10 in H95:
$$V = -2.5~\log ~(DN/t) + 0.254(V-I) + 0.012(V-I)^2 + $$
$$22.093 + 2.5\log(GR) + 0.1 + 0.05.$$
Here GR is gain ratio as defined in H95, the term 0.1 corrects the formula for 
infinite aperture and 0.05 is the correction due to long exposures. 
$V-I$ was calculated iteratively. Luminosity flux was corrected for 
the cosmological distortion by a factor $(1+z)^4.$

Luminosities have been corrected for absorption in our Galaxy according
to Schlegel et al. (\cite{schl}).

\begin{figure*}
\begin{center}
\setlength{\unitlength}{1cm}
\begin{picture}(32,6)
\put(-0.5,5.5){\includegraphics{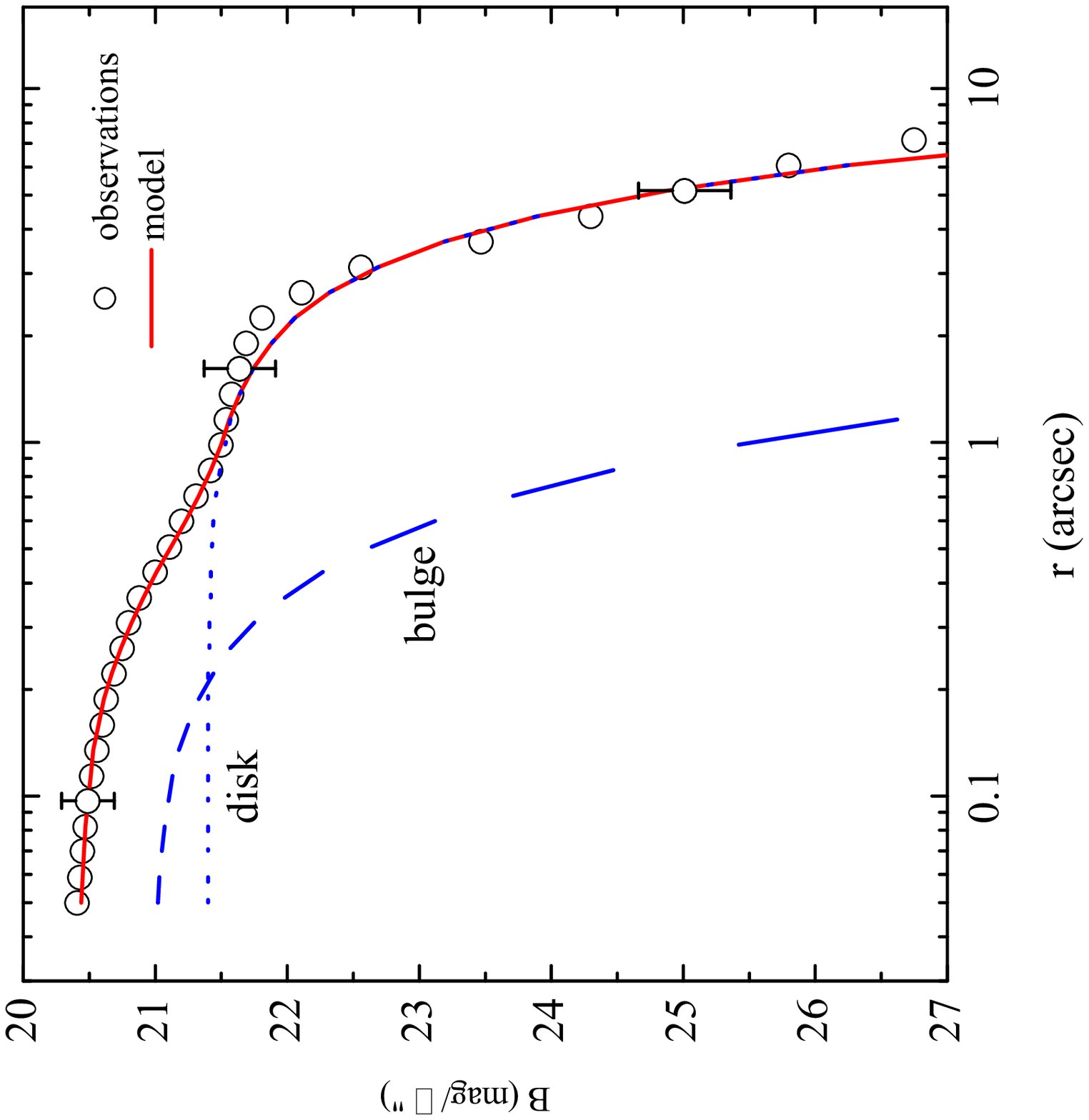}}
\put(5.5,5.5){\includegraphics{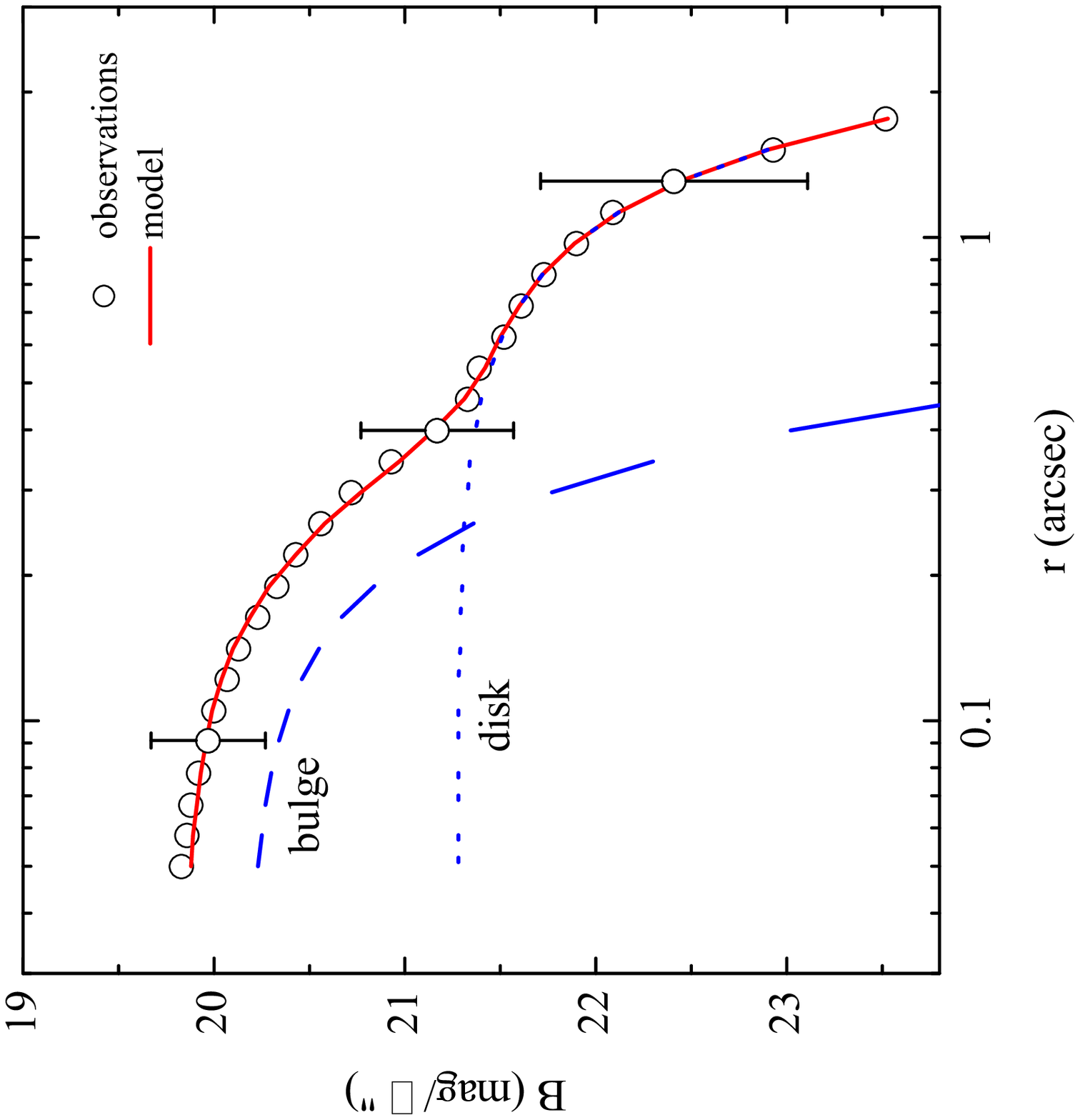}}
\put(11.5,5.5){\includegraphics{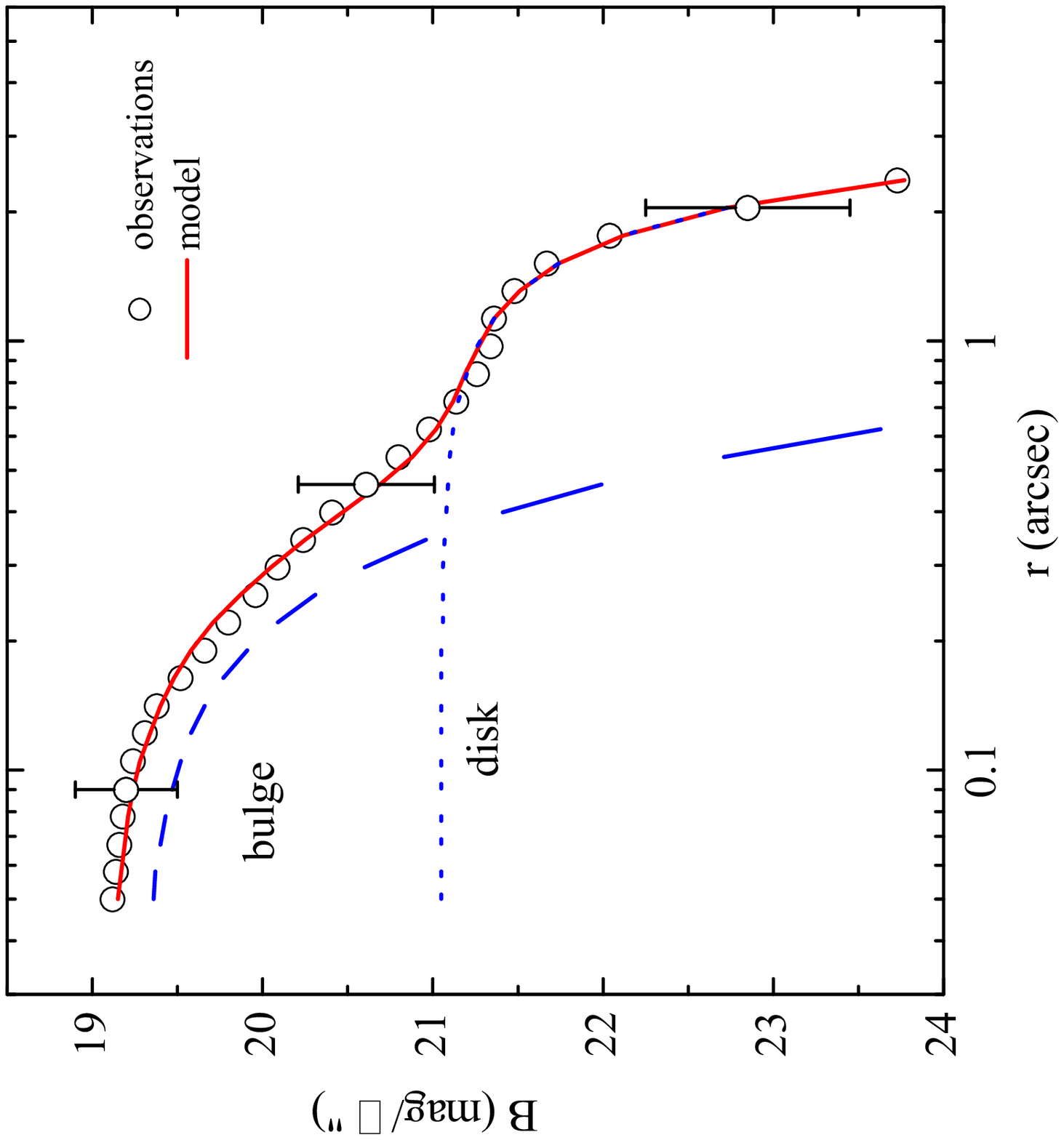}}
\end{picture}
\end{center}
\caption{Surface brightness distribution of the galaxies 074-2237, 
         064-4412, and 094-2210 (from left to right) in rest-frame B color 
         (open circles). Solid line -- surface brightness distribution 
         of the best fit model, dashed lines -- surface brightness 
         distribution of model components.}
\label{g123b}
\end{figure*}

We estimated the k-correction in two ways. First, the observed colors 
were transformed to such standard colors for which the k-correction 
would be minimal. For this we used the method described in van Dokkum 
\& Franx (\cite{do:fr1}) and Kelson et al. (\cite{kels2}). At redshift 
$z=0.15$ a suitable transformation would be from V to B and for redshifts 
0.9 and 0.99 from I to B. For all three galaxies we used synthetic 
spectra of a Sbc--Scd galaxy derived by Coleman et al. (\cite{cole}). 
Zero points relating AB magnitudes and Johnson magnitudes were taken 
from Frei \& Gunn (\cite{fr:gu}). Second, the k-corrections were 
estimated by using the chemical evolution models by Schulz et al. 
(\cite{schu}). Again, at $z=0.15$ V color was transformed to B color, 
assuming the rest-frame $\rm (B-V) = 0.63$, at redshifts 0.90 and 0.99 
I color was transformed to B color, assuming the rest-frame $\rm (B-I) 
= 1.76$. For the nearest galaxy, we took solar metallicity, for the two
others $Z=0.008$. The results obtained by two methods nearly coincide in
the case of the nearest galaxy, for more distant galaxies the agreement 
is within $\pm 0.15^{\rm m}$. Averaging the results of the two methods 
yields the following relations:
$$\begin{array}{ll}
B(z=0.15) & = V + 0.32,\\
B(z=0.90) & = I + 0.90,\\
B(z=0.99) & = I + 1.02.\\
\end{array} $$
Here $V$, $I$ and $(V-I)$ are observed Johnson magnitudes and $B$ is the 
Johnson magnitude in galactic rest-frame. The final galactic surface
brightness profiles are presented in Fig.~\ref{g123b} by open circles.
Error bars characterize uncertainties of both background determination
and isophote construction.

\section{Galactic models}
Although several stellar populations can be discriminated in spiral 
galaxies, in the case of modeling very distant galaxies it is reasonable to
limit the main stellar components to two -- the bulge and the disk.
Parameters of any other stellar component (central nucleus, metal poor
halo, thick disk) could not be determined at present from available
photometric observations. To construct a dynamical model, a dark
matter component -- the dark halo must be added to visible components.
Of course, the photometrical and dynamical models must be mutually consistent.

When handling only surface photometry data, it is convenient to describe
the galactic structure with the help of some simple parametric formulae for
the surface density. Most common formulae are the de Vaucouleurs formula, the
exponential law or the more general S\'ersic formula. Most flexible is the
S\'ersic formula with an additional parameter, allowing to control the
steepness of the surface density decrease (S\'ersic \cite{sersic}). Density
distribution parameters must be determined e.g. by the least squares method.
Using also kinematic data and constructing a dynamical model
consistent with the photometry, the same density distribution law must be
used for rotation curve modeling (and for velocity dispersion curve, if
possible). In the case of the S\'ersic formula, it can be done for spherical
systems with an integer S\'ersic index $m$ (Mazure \& Capelato \cite{ma:ca}).
For a noninteger index and ellipsoidal surface density distribution a consistent
solution for rotation curve calculations is not known.

In the present paper, the density distribution parameters are determined
by the least squares method and they can have any value. Thus, it is not reasonable
to use the S\'ersic law for surface density distribution. It is more physical
to start from the spatial density law. An analytical expression for spatial 
densities, reproducing well the S\'ersic formula, was derived by Trujillo
et al. (\cite{truj2}). Although this formula is useful, we decided to use
for the bulge and for the disk an analyticaly simpler form (\ref{eq1}), 
allowing to use more easily least-square fitting simultaneously for light
distribution and rotation curve. 

In our model the visible part of a galaxy
is given as a superposition of the bulge and the disk. The spatial density
distribution of every visible component (the bulge and the disk) is approximated 
by an inhomogeneous
ellipsoid of rotational symmetry with constant axial ratio $q$ and
the density distribution law (Einasto \cite{einasto})
\begin{equation}
\rho (a)=\rho (0)\exp [ -( a/(ka_0))^{1/N} ] ,
\label{eq1}
\end{equation}
where $\rho (0)=hM/(4\pi q a_0^3)$ is the central density, $M$ is the 
component mass. Next, $a= \sqrt{R^2+z^2/q^2}$, where $R$ and $z$ are 
two cylindrical coordinates. $a_0$ is the harmonic mean radius
\begin{equation}
a_0^{-1} = \int_0^{\infty} a^{-1} dM ,
\end{equation}
where $dM = 4\pi q a^2\rho (a) da$ is the mass of the ellipsoidal
shell of thickness $da$. Coefficients $h$ and $k$ are normalizing parameters, depending
on $N$, which allows to vary the density behavior with $a$. The harmonic
mean radius characterizes rather well the real extent of a component,
independently of the parameter $N$. The definition of normalizing
parameters $h$ and $k$ and their calculation is described in Tenjes et
al. (\cite{tenj1}), Appendix B. Eq.~\ref{eq1} allows sufficiently precise
numerical integration and has a minimum number of free parameters.

The dark matter (DM) distribution is represented by a spherical isothermal
law
\begin{equation}
\rho (a) = \cases{\rho (0) ([1+({a\over a_c})^2]^{-1} -
                      [1+({a^0\over a_c})^2]^{-1}) & $a \leq a^0$
                      \cr
                    0         &      $a>a^0$.    \cr }
\label{eq3}
\end{equation}
Here $a^0$ is the outer cutoff radius of the isothermal sphere.

The density distributions for the bulge and the disk were projected
along the line of sight, divided by their mass-to-luminosity ratios
$f$ and their sum gives us the surface brightness distribution of
the model
\begin{equation}
L(A) = 2 \sum_{i=1}^2 {q_i\over Q_i f_i} \int_A^{\infty}
          {\rho_i (a) a~da\over (a^2 - A^2)^{1/2}} ,
\label{eq4}
\end{equation}
where $A$ is the major semiaxis of the equidensity ellipse of the
projected light distribution, $Q_i$ are their apparent axial ratios
$Q^2=\cos^2\gamma+q^2\sin^2\gamma$. The angle between the plane of a
galaxy and the plane of the sky is denoted by $\gamma$. Summation
index $i$ designates two visible components, the bulge and the disk.

The masses of the components we determined from the rotation law
\begin{equation}
v_i^2(R) = 4\pi q_i G \int_{0}^{R} {\rho_i (a) a^2 da\over
              (R^2-e_i^2a^2)^{1/2}},
\label{eq5}
\end{equation}
\begin{equation}
   V^2(R) = \sum_{i=1}^3 v_i^2(R),
\label{eq6}
\end{equation}
where $G$ is the gravitational constant, $e=\sqrt{1-q^2}$
is eccentricity, and $R$ is the distance in the equatorial plane of
the galaxy. Summation is over all three components now.

Model parameters $a_0$, $q$, $L_B$, $M$ and $N$ for the bulge and
the disk were determined by a subsequent least-squares approximation
process. As the first step, we made a crude estimation of the
population parameters. The purpose of this step is only to avoid
obviously unphysical parameters -- relations (\ref{eq4}) and (\ref{eq5}) are
nonlinear and fitting the model to the observations is not a
straightforward procedure. Next a mathematically correct solution
for each galaxy was found. Details of the least squares approximation
and the general modeling procedure were described by Einasto \& Haud 
(\cite{ei:ha}), Tenjes et al. (\cite{tenj1}, \cite{tenj2}). 

The rotation curves of the galaxies 074-2237, 064-4412, 094-2210 have
a rather small extent (see Figs. \ref{g1v1}--\ref{g3v1}). We do not have
any additional information about the dynamics of these galaxies at
large galactocentric radii. For this reason, the outer cutoff radius of
the DM component remains undetermined at present. We fixed
$a^0 = 5 a_0$ on the basis of the structure of nearby galaxies. This
value influences the rotation curve behavior only in extreme outer regions 
of galaxies and does not influence our results. Then we assumed by
analogy with nearby galaxies that the rotation curve of these galaxies
remains flat at least up to 30 kpc (Sofue \& Rubin \cite{so:ru}).

As we have no information about velocity dispersions in the
central regions of our galaxies we can calculate only the circular
velocities (Eqs.~\ref{eq5}--\ref{eq6}). If the value of velocity
dispersions is comparable to the value of rotational velocity at
a particular distance, the latter does not correspond to the value
of circular velocity. The difference is known in galactic dynamics
as the asymmetric drift. Typical emission-line dispersions in galaxies
at intermediate redshifts are 30--100 km/s (Koo et al. \cite{koo};
Guzman et al. \cite{guzm}; Im et al. \cite{im1}). Thus, within central
$0.5\arcsec - 0.7\arcsec$ model velocities must be higher than the
observed rotational velocities.

\section{Results and discussion}

Calculated from our photometry, the total apparent Johnson V luminosity
of the galaxy 074-2237 is $V = 19.28\pm 0.05$, which coincides well
with the $\rm V_{\rm 606}$ luminosity estimates by Vogt et al. (\cite{vogt1})
18.70 and Simard et al. (\cite{sima}) 18.78. The mean apparent color
index is $\rm (V-I) = 1.5\pm 0.2$. The difference between the Johnson V and
$\rm V_{606}$ is $0.49^m$, if $\rm (V-I) =$1.5.

The calculated total apparent Johnson I luminosity of the galaxy
064-4412 is $I = 22.43$, which also coincides well with the $I_{\rm 814}$
luminosity estimates by Vogt et al. (\cite{vogt1}) 22.38 and by Simard
et al. (\cite{sima}) 22.33. The mean apparent color index is $\rm (V-I) = 2.2$.
The difference between the Johnson I and $\rm I_{814}$ is $0.12^m$ if $\rm (V-I)
=$2.3.

The total apparent Johnson I luminosity of the galaxy 094-2210 from our
photometry is $I = 21.47$. The $I_{\rm 814}$ luminosity estimate by Vogt
et al. (\cite{vogt1}) is 21.41 and by Simard et al. (\cite{sima}) it is 21.21.
The mean apparent color index is $\rm (V-I) = 2.4$.

\begin{table*}
   \caption[]{Galactic model parameters.}
   \label{tab2}
\begin{tabular}{llllllllll}
\hline
Name &$q$   &$a_0$&$M$&$M/L_B$&$N$& $h$  & $k$    & $L_B$ & $M^a$\\
     &          &(kpc)&$\mathrm 10^{10}M_{\sun}$& &  &   & $\mathrm 10^{10} L_{\sun}$ &
     $\mathrm 10^{10} M_{\sun}$ \\
\hline
074-2237 &  &     &     &     &      &       &        &       &      \\
bulge & 0.7 & 1.1 & (0.2) & 2.2 & 0.75 & 2.459 & 0.7822 & 0.092 & (0.05) \\
disk  & 0.1 & 7.1 & 6.4 & 4.4 & 0.50 & 1.571 & 1.128  & 1.46  & 6.5  \\
dark matter& & 60.& (170.)&     &      & 14.82 & 0.1511 &       & (170.) \\
064-4412 &  &     &     &     &      &       &        &       &      \\
bulge & 0.7 & 1.5 & (0.3) & 0.78 & 0.51 & 1.592 & 1.117 & 0.38  & (0.05) \\
disk  & 0.1 & 7.3 & 2.7 & 1.17 & 0.43 & 1.397 & 1.229  & 2.30  & 3.0  \\
dark matter& & 40.& (150.)&     &      & 14.82 & 0.1511 &       & (150.) \\
094-2210 &  &     &     &     &      &       &        &       &      \\
bulge & 0.7 & 1.7 & (2.1) & 1.9 & 0.67 & 2.115 & 0.8801 & 1.08  & (0.7)  \\
disk  & 0.1 & 9.3 & 6.4 & 1.19 & 0.25 & 1.080 & 1.445  & 5.38  & 8.0  \\
dark matter& & 60.& (250.)&     &      & 14.82 & 0.1511 &       & (250.) \\
\hline
\end{tabular}
\begin{list}{}{}
\item[$^a$] Mass distribution model with low bulge masses (upper panels
in Figs~4--6).\\
\end{list}
\end{table*}

The final two-component models fit the photometric B-profiles with a mean
relative error of 0.2--0.5 per cent. For the fitting of the rotation curves,
we constructed two models for each galaxy. One kinematical model
corresponds to the mathematically best fit at all distances, even in
the central regions. However, due to the arguments given in the last
paragraph of the previous section, this model is unphysical, because
it underestimates the bulge mass. Therefore, a second model was
constructed with the bulge mass fixed at higher values. At
present we have no independent basis to estimate the bulge mass.
We hope that two different models allow the reader to understand
how the bulge mass variation influences the overall behaviour of
rotation curves in all three concrete cases. Real bulge masses and 
thus also mass-to-light ratios may be higher than the values resulting 
from Table~\ref{tab2}. To derive real bulge mass-to-light ratios, 
additional observations (central stellar velocity dispersions) are 
needed.

\begin{figure}
\resizebox{\hsize}{!}{\includegraphics*{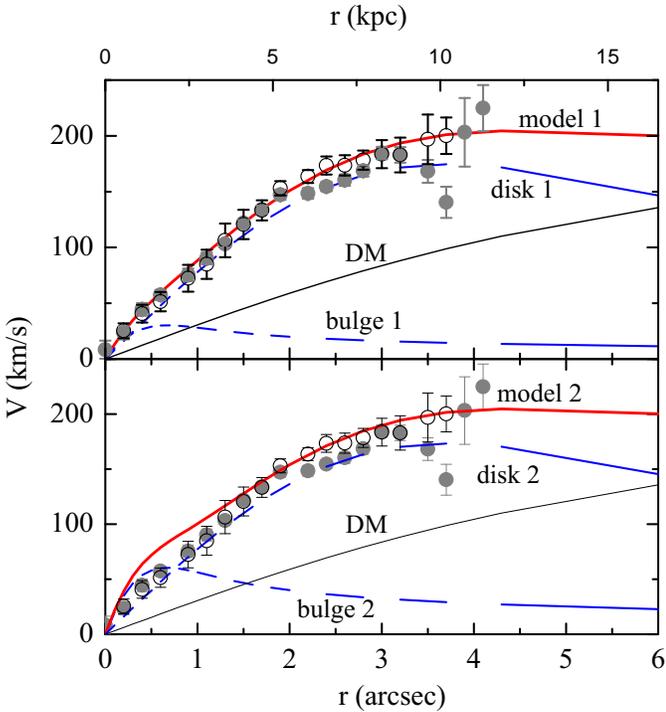}}
    \caption{Rotation curve of the galaxy 074-2237. Open and grey circles --
     observations by Vogt et al. (1996) at the opposite sides of the galaxy,
     thick solid line -- rotation curve from the mass model, dashed lines
     -- rotation velocities due to visible components, thin solid line --
     rotation velocities due to the dark matter component. Upper panel
     (model 1) corresponds to low bulge mass, lower panel (model 2)
     corresponds to higher bulge mass.}
   \label{g1v1}
\end{figure}

\begin{figure}
\resizebox{\hsize}{!}{\includegraphics*{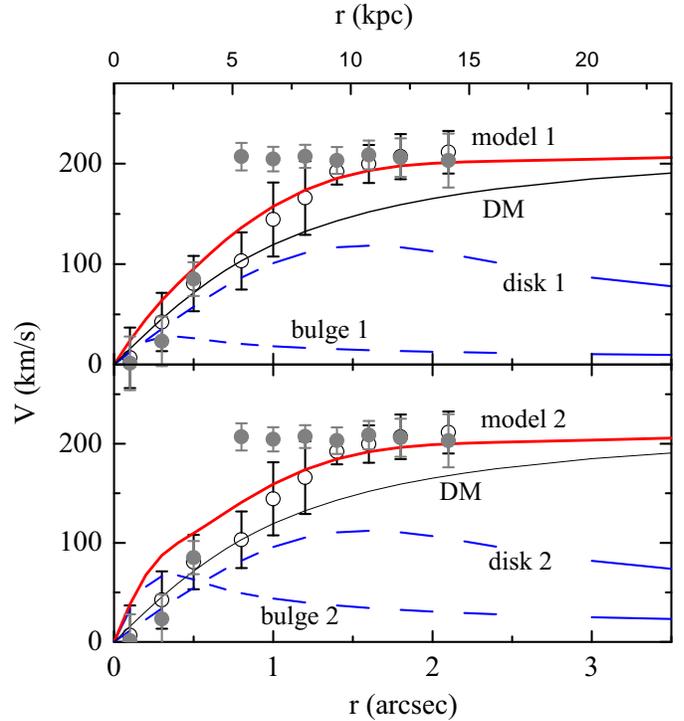}}
   \caption{Rotation curve of the galaxy 064-4412. Designations are the same
   as in Fig.~\ref{g1v1}.}
   \label{g2v1}
\end{figure}

\begin{figure}
\resizebox{\hsize}{!}{\includegraphics*{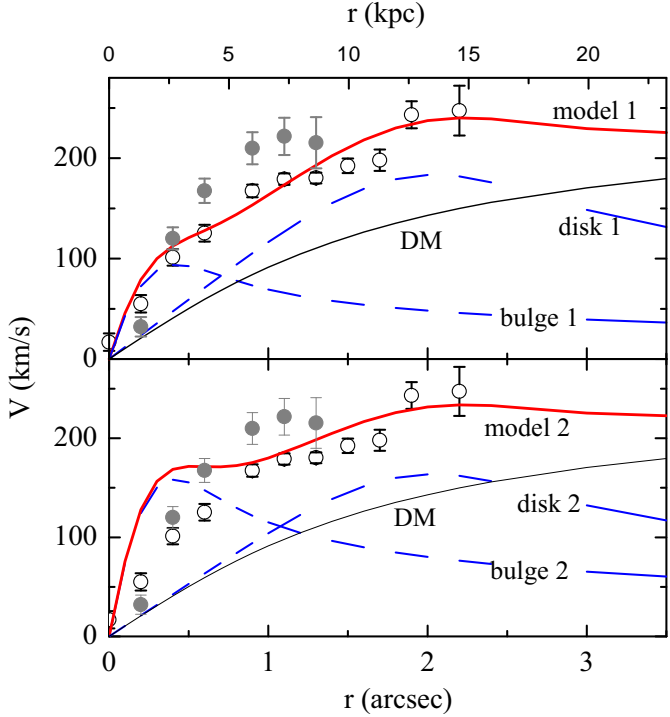}}
   \caption{Rotation curve of the galaxy 094-2210. Designations are the same
   as in Fig.~\ref{g1v1}.}
   \label{g3v1}
\end{figure}

The parameters of the final models (the axis ratio $q$, the harmonic 
mean radius $a_0$, the total mass of the population $M$, the structural 
parameter $N$, the dimensionless normalizing constants $h$ and $k$, 
B-luminosities and corresponding mass-to-light ratios) are given in 
Table~\ref{tab2}. Bulge and DM masses are given in parentheses 
to indicate that on the basis of present observational data it is not 
possible to determine these values uniquely. The last column in 
Table~\ref{tab2} gives the masses for low bulge mass models. The final 
models are denoted by thick solid lines in Figs.~\ref{g123b}--\ref{g3v1}.
Models of individual visible components are represented by dashed lines,
the model of the dark matter component by a thin solid line. In Table 
\ref{tab3}, calculated from the models, integrated parameters are given: 
apparent I-band magnitude, total mass of the visible matter, total 
intrinsic luminosity and mass-to-light ratio of the visible matter. 

\begin{table}
   \caption[]{Derived integrated parameters from models.}
   \label{tab3}
\begin{tabular}{lllll}
\hline
Name$^a$ & $m_I$ & $M_{\rm vis}$         & $L_B$                 & $(M/L)_{\rm vis}$ \\
         & (mag) & $\rm 10^{10}M_{\sun}$ & $\rm 10^{10}L_{\sun}$ &                   \\
\hline
074-2237 & 17.78 & 6.6                   & 1.55                  & 4.3               \\
064-4412 & 22.43 & 3.0                   & 2.68                  & 1.1               \\
094-2210 & 21.47 & 8.4                   & 6.46                  & 1.3               \\
\hline
\end{tabular}
\end{table}

It is interesting to compare the central surface brightnesses of galactic
disks with the results obtained for local galaxies. For the galaxies
modeled here, the central brightnesses of disks are $\mu_B(0) =$ 21.4,
21.3, 21.1. The mean value for two galaxies at redshifts $z=0.9-1$ is
$\mu_B(0) = 21.2$. For local galaxies the original value by Freeman
(\cite{freeman}) was 21.65. However, this value probably depends on the
galactic morphological type and luminosity, and thus has considerable
scatter (Graham \& de Blok \cite{gr:bl}). For galaxies at redshifts $z=
0.5-1$ Schade et al. (\cite{scha2}) derived $\mu_B(0)=20.0$, when 
transformed to conventional broadband magnitudes, but also
with a rather significant scatter. Our values are clearly higher.
However, it must be taken into account that central brightnesses of 
disks have been determined by Schade et al. (\cite{scha2}) within the assumption
of exponential disks and $r^{1/4}$ bulge. When fitting the disks of our two 
higher redshift galaxies with parameters $N=1$, the resulting disk central
brightnesses are brighter by 0.6 mag, i.e. the mean value is $\mu_B(0) 
= 20.6$. Disk central brightness is sensitive also to bulge parameters.
Our bulge profile is quite different from the $r^{1/4}$ law. And finally,
when taking into account also the uncertainties in k-correction,
our values for the disk central brightnesses are not peculiar. 

Figure~\ref{g123b} shows that the light distribution in outer parts of the
galaxies studied in the present paper is rather steep (the parameter $N$ is
small). In all cases, the disk density distribution in outer parts is steeper 
than for a simple exponential disk. Central concentration parameters (Trujillo 
et al. \cite{truj1}) for these galaxies are rather small, about 0.15.
Steep decrease of the surface brightness in outer parts of galaxies may
be the result of the outer truncation of the disk. According to Pohlen et al.
(\cite{pohl}), the truncation occurs at distances of about 3 disk scale lengths.
In the case of galaxies studied here, the surface photometry does not extend to
so large distances.

The central density of the dark matter $\rho (0) = hM/(4\pi a_0^3),$
characterizing the rotation velocity gradient at the center caused by
the dark halo, is rather well determined from observations and nearly
independent of the choice of the dark halo outer cutoff radius.
Calculated from our models, the central densities of DM
components are $\rho (0) =$ 0.0092, 0.028, and 0.015 in units $\rm M_{\sun} /
pc^3$. It is important to stress that these values are calculated for maximum
disk models and are thus the lower limits. This must be taken into
account especially in the case of the first value (the galaxy 074-2237)
where models with lower disk mass and higher DM density give equally
good fit for the rotation curve. The visible and DM
masses become equal for this galaxy at $R=17$~kpc in our maximum disk model.
A mass distribution model of the galaxy 074-2237 was conctructed also by
Fuchs et al. (\cite{fuch}) who used for mass determination a model of spiral
arm formation via swing-amplification. They concluded that in order to create
a two-armed spiral structure, higher DM densities are needed when
compared with the maximum disk models. According to their model, the central
DM density of 074-2237 is 0.025.

\begin{figure}
\resizebox{\hsize}{!}{\includegraphics*{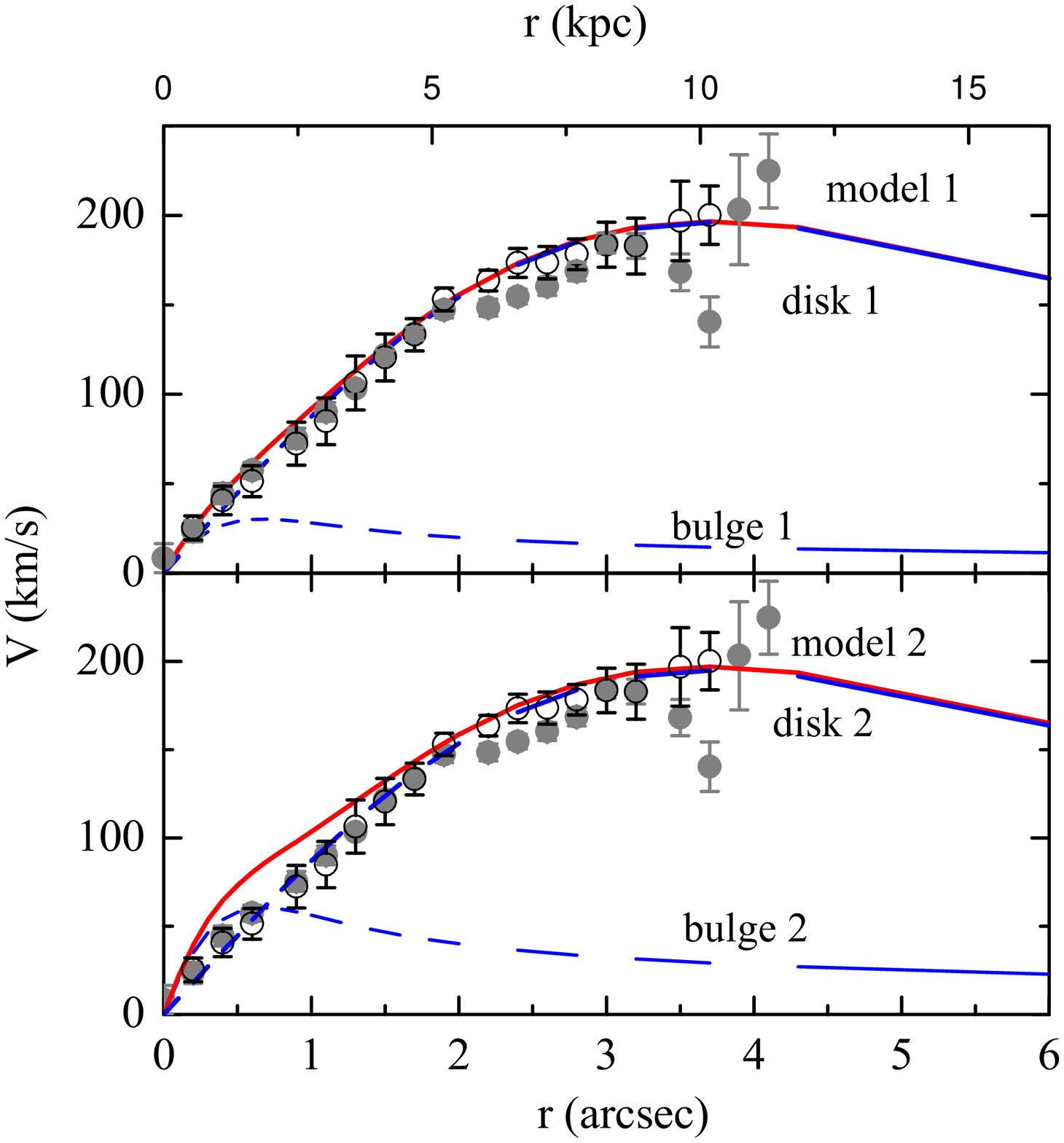}}
    \caption{Rotation curve model of the galaxy 074-2237 without the dark matter
     component. Disk masses are 8.2 (disk 1) and 8.1 (disk 2) in units of
     $\rm 10^{10}M_{\sun}$. Designations are the same as in Fig.~\ref{g1v1}.}
   \label{g1v2}
\end{figure}

\begin{figure}
\resizebox{\hsize}{!}{\includegraphics*{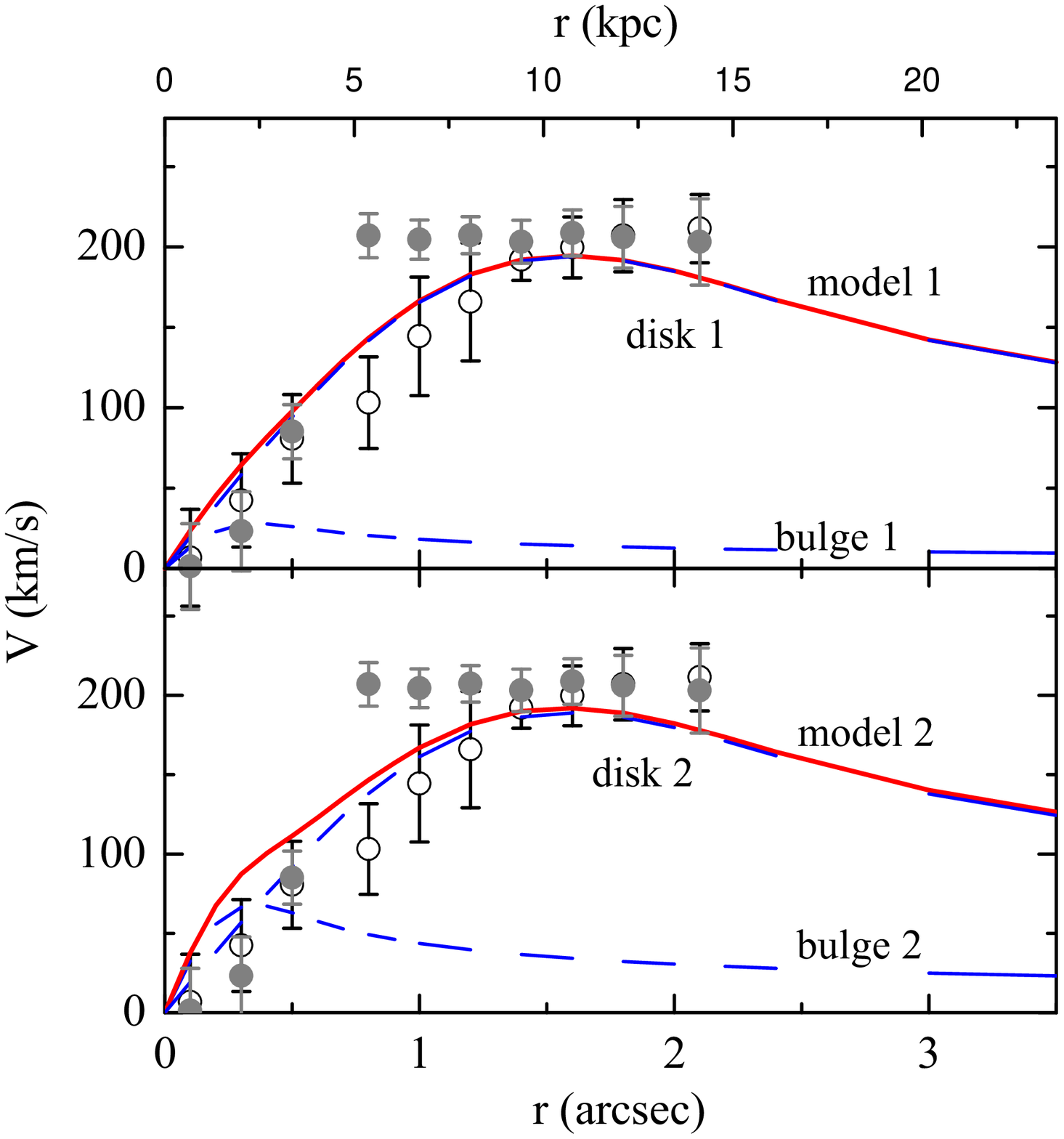}}
   \caption{Rotation curve model of the galaxy 064-4412 without the dark matter
   component. Disk masses are 8.1 (disk 1) and 7.7 (disk 2) in units of
   $\rm 10^{10}M_{\sun}$. Designations are the same as in Fig.~\ref{g1v1}.}
   \label{g2v2}
\end{figure}

\begin{figure}
\resizebox{\hsize}{!}{\includegraphics*{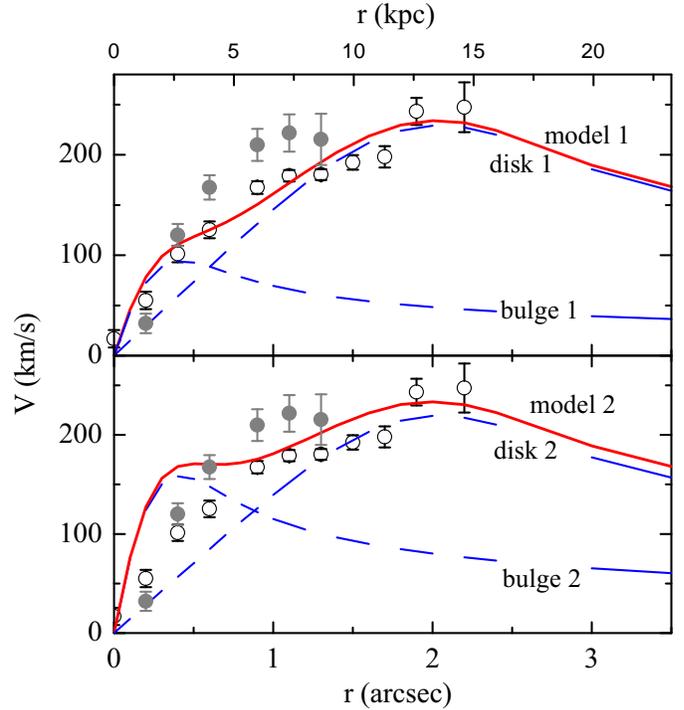}}
   \caption{Rotation curve model of the galaxy 094-2210 without the dark matter
   component. Disk masses are 12.5 (disk 1) and 11.4 (disk 2) in units of
   $\rm 10^{10}M_{\sun}$. Designations are the same as in Fig.~\ref{g1v1}.}
   \label{g3v2}
\end{figure}

It is possible to construct mass distribution models for galaxies
without assuming the existence of a DM component. 
Corresponding models are given in Figs~\ref{g1v2}--\ref{g3v2}. It is seen
that the calculated rotation curves start to decrease after the last
observed point of the rotation curves. Thus, within the available at present
observations, the models without any dark halo fit well with observations.

As mentioned earlier, due to the limitied extent of rotational 
curves it is not possible to determine any parameters of the DM 
distribution. Figures~\ref{g1v2}--\ref{g3v2} show that rotation curves 
can be well fitted without DM. However, we think that there are 
enough independent arguments in support of the existence of DM
and thus, realistic models must include a DM component. For this 
reason, on the basis of nearby galaxies weumed that rotation curves
remain flat up to the distances of about 30~kpc (see Sofue \& Rubin 
\cite{so:ru}). This assumption still does not enable to determine the 
total DM mass and its characteristic radius. Only the central density 
of the DM component can be found. Therefore, the mass and the radius 
of the DM component given in Table~\ref{tab2} are simply fixed. It is 
possible to fix, e. g. the DM mass two times smaller, decreasing 
correspondingly the DM radius in a way that the central density remain 
unchanged. The central density of DM can be dermined from models within 
the assumption of the flatness of rotation curves. The central density 
of DM is not sensitive to the uncertainties of bulge mass.

The central density of the dark matter distribution is a parameter 
allowing to constrain possible cosmological models and corresponding 
parameters at early periods of structure formation (see eg. Navarro \& Steinmetz 
\cite{na:st}, Shapiro \& Iliev \cite{sh:il}, Khairul Alam et al. \cite{khai},
Zentner \& Bullock \cite{ze:bu}). We think it is reasonable to derive also the
central DM densities at different redshifts. In the present paper we have used
only one density distribution model for DM  (Equation \ref{eq3}). In
subsequent papers we intend to enlarge the sample of modeled galaxies 
and study the dependence of model parameters on different DM distribution
models (e.g. models by Navarro et al. \cite{nava}, Burkert \cite{burkert}).

\begin{acknowledgements}
We would like to thank the anonymous referee for useful comments and
suggestions. We thank Dr. U. Haud for making available his
programs for light distribution model calculations. Surface
photometry data used in the present study are from STScI Archive and
we thank the members of observational proposals 5109 (PI J.
Westphal) and 5090 (PI E. Groth) for their work. We acknowledge
the financial support from the Estonian Science Foundation (grant
4702) and DAAD (grant A0209036). Part of the paper was written at 
Goettingen University Observatory and we thank the staff for 
hospitality. We also thank Dr. N. Vogt for communicating his  
data in numerical form as electronic files.
\end{acknowledgements}

\end{document}